\RequirePackage{ifpdf}
\documentclass[12pt,letterpaper]{JHEP3}         
\usepackage{amssymb,amsfonts}

\usepackage{wrapfig}
\usepackage{epsfig}

\def\be{\begin{eqnarray}}
\def\ee{\end{eqnarray}}
\newcommand{\nn}{\nonumber}
\newcommand\para{\paragraph{}}

\newcommand{\eqn}[1]{(\ref{#1})}

\def\Dslash{\,\,{\raise.15ex\hbox{/}\mkern-12mu D}}
\def\Dbarslash{\,\,{\raise.15ex\hbox{/}\mkern-12mu {\bar D}}}
\def\delslash{\,\,{\raise.15ex\hbox{/}\mkern-9mu \partial}}
\def\delbarslash{\,\,{\raise.15ex\hbox{/}\mkern-9mu {\bar\partial}}}
\def\pslash{\,\,{\raise.15ex\hbox{/}\mkern-9mu p}}
\def\calDslash{\,\,{\raise.15ex\hbox{/}\mkern-12mu {\cal D}}}

\def\implies{\Rightarrow}

\def\lae{\mathrel{\mathop{\smash{\lower .5 ex \hbox{$\stackrel<\sim$}}}}}
\def\lae{\mathrel{\mathop{\smash{\lower .5 ex \hbox{$\stackrel>\sim$}}}}}

\title{Vortices and Monopoles in a Harmonic Trap}

\author{David Tong and Carl Turner\\
Department of Applied Mathematics and Theoretical Physics, \\
University of Cambridge, \\ 
Cambridge, CB3 OWA, UK \\
{\tt  d.tong, c.turner@damtp.cam.ac.uk}\\}

\abstract{The $\Omega$-deformation is a harmonic trap, penning certain excitations near the origin in a manner consistent with supersymmetry. Here we explore the dynamics of BPS monopoles and vortices in such a trap. We pay particular attention to monopoles in the Higgs phase, when they are confined to a vortex string. Unusually for BPS solitons, the mass of these confined monopoles is quadratic in the topological charges. We compute an index theorem to determine the number of collective coordinates of confined monopoles. Despite being restricted to move on a line, we find that they have a rich dynamics. As the strength of the trap increases, the number of collective coordinates can change, sometimes with constituent monopoles disappearing, sometimes with new ones emerging.
 }

\begin{document}
\pagestyle{plain} \setcounter{page}{1}
\newcounter{bean}
\baselineskip16pt \setcounter{section}{0}

\section{Introduction}

The $\Omega$-background is a deformation of supersymmetric gauge theories which breaks translational invariance. It has proven to be both a powerful tool for computation \cite{nikitaisitcold,inyourlittlecorner}, and a useful device to highlight connections between different theories, most notably four-dimensional gauge theories and two dimensional integrable systems \cite{oftheglobe,nekwitten}. 

\para
In this paper we take a more prosaic view of the $\Omega$-background. We view it simply as a harmonic trap, analogous to those which arise in condensed matter physics. Its role is to restrict certain excitations to lie close to the origin. The excitations that we will be interested in are solitons. In supersymmetric gauge theories, BPS solitons typically have a number of nice properties, both physical and mathematical. The $\Omega$-background provides a harmonic trap which is consistent with supersymmetry and, correspondingly, preserves many of these nice properties. 

\para
There are at least two motivations to study solitons in the $\Omega$-background. The first is purely classical. A harmonic trap squeezes solitons together. Yet this is often where solitons are at their most interesting. They no longer appear as a point-like objects and their extended, non-linear nature becomes apparent. They lose their individuality, merging into each other to form something new, often with interesting structures and collective excitations. 

\para 
The second motivation is more quantum in origin. Solitons provide a semi-classical springboard to study some of the interesting dualities that are induced by the $\Omega$-background. In particular, we have in mind the 4d/2d duality described in \cite{nick1,nick2,mina}, relating the Seiberg-Witten curve to twisted superpotentials of 2d sigma-models. This is an extension of an earlier duality \cite{nick3,nick4} which found an explanation in the dynamics of vortex strings \cite{shifyung,meami2}. Here we study the vortex strings relevant for the extended duality. More recently, there have been studies of  3d gauge theories \cite{bull} and 5d gauge theories \cite{soojong} in the presence of the $\Omega$-background  and we will describe the vortices and monopoles relevant for these theories.

\subsubsection*{What We Do}

We study solitons in ${\cal N}=2$ four-dimensional theories with an $\Omega$-deformation in a single plane. We are not the first to study solitons in this background. A number of properties of monopoles were explored in \cite{ito1,ito2} and a range of other solitons were described in \cite{hengyu}. 

\para
In this paper, we will describe three different types of solitons: monopoles, vortices and confined
 monopoles. Our interest is primarily in the latter. These are monopoles in the Higgs phase, where they  appear as beads threaded on vortex strings, yet remain BPS \cite{monohiggs}.   Our main results are the formula \eqn{monomass} for the mass of the confined monopole in a trap, and the index theorem \eqn{index} for these objects.

\para
We find that the presence of the harmonic trap endows these confined monopoles with a rich dynamics.  A generic, higher-charge monopole can split into constituent monopoles, each free to move up and down along the string. However, the mass of each of these constituents has an extra term  which, unusually for BPS solitons, is quadratic in the magnetic flux charges. This can be thought of a binding energy between the monopole and other flux tubes which also lie in the trap. 

\para
As one increases the strength of the harmonic trap, the number of collective coordinates jumps. Sometimes this reflects the fact that some of the constituent monopoles  become massless and disappear; sometimes it reflects the fact the new constituent monopoles appear. One of the surprising features is that monopoles with charges that one might naively have thought of as anti-BPS can apparently become BPS in the presence of the trap. Much of the paper is devoted to telling this story. 

\para
The paper also includes a number of other results. In particular, in two appendices we study the dynamics of vortices in the presence of a harmonic trap. The effect of the trap is to induce a potential on the vortex moduli space, so that the ground state of vortices is an incompressible disc lying at the origin of the plane. We show that, for $U(1)$ vortices,  the collective excitations of this disc have a description as a field theory living on the edge of the disc.

\section{Solitons in a Harmonic Trap}

The theory of interest consists of  a $U(N)$ gauge field $A_\mu$, coupled to a real adjoint scalar $\phi$ and $N_f$ fundamental scalars $q_i$.  

\para
We impose on these fields an external, harmonic trap whose strength is parameterised by $\omega$. This is the $\Omega$-deformation.  The trap breaks translational symmetry, penning certain excitations close to the origin in the $(x^1,x^2)$ plane; however they remain free to move in the $x^3$ direction. This is the form of the $\Omega$-deformation discussed in \cite{oftheglobe}.

\para
The Lagrangian for this supersymmetric trap was first derived in \cite{shadchin}. (The reference \cite{hengyu} contains a useful review of this work.) It is given by
\be {\cal L} &=& - \frac{1}{4e^2} {\rm Tr}\,F_{\mu\nu}F^{\mu\nu} - \frac{1}{2e^2} {\rm Tr}\,({\cal D}_0 \phi  - \omega(x^2 E_1 - x^1 E_2))^2 -  \frac{1}{2e^2} {\rm Tr}\, ({\cal D}_1\phi - \omega x^1 B_3)^2\nn\\   &&
-  \,\frac{1}{2e^2} {\rm Tr}\, ({\cal D}_2\phi - \omega x^2 B_3)^2 - \frac{1}{2e^2} {\rm Tr}\,({\cal D}_3 \phi  + \omega(x^1 B_1 + x^2 B_2))^2 + \sum_{i=1}^{N_f} |{\cal D}_\mu q_i|^2
\nn\\ &&  +\, \frac{e^2}{2}{\rm Tr}\,( \sum_i q_i q_i^\dagger - v^2 )^2 + \sum_{i=1}^{N_f} |(\phi- m_i) q_i - i\omega (x^1 {\cal D}_2 q_i + x^2 {\cal D}_1 q_i)|^2 
\label{lag}\ee
Here ${\cal D}\phi = \partial \phi - i[A,\phi]$ while ${\cal D} q = \partial q - i Aq$.  The electric field is $E_i = F_{0i}$ and the magnetic field is $B_i = \frac{1}{2}\epsilon_{ijk} F_{jk}$. In this paper, we focus on the theory with $N_f=N$; this is the minimal theory admitting vortices.

\para
The Lagrangian  admits a completion to a theory with ${\cal N}=2$ supersymmetry. In addition to the  fermions, the theory with ${\cal N}=2$ supersymmetry has a further real adjoint scalar in the  vector multiplet and $N_f$ anti-fundamental  scalars in the hypermultiplets. There are non-trivial terms in the $\Omega$-deformation involving all  these fields \cite{shadchin}. (In particular, there is an extra contribution to the D-term involving the imaginary part of the adjoint scalar.)   However, it turns out that these extra bosonic fields vanish on our soliton backgrounds and, to avoid cluttering equations, we have chosen to omit them from the start.

\para
The Lagrangian \eqn{lag} admits a number of different soliton solutions depending on  the values of  $v^2$ and the real-valued masses $m_i$. We now review these different solutions.

\subsection{Monopoles}\label{monosec}

When $v^2=0$, we can turn off the fundamental scalars, $q_i=0$. There is no potential for the adjoint scalar $\phi$ and we are free to specify a vacuum expectation value that lies in the Cartan sub-algebra. This breaks the gauge group $U(N) \rightarrow U(1)^N$.
%
%

\para
With these boundary conditions, the theory admits BPS magnetic monopoles. These obey the deformed Bogomolnyi equations \cite{ito1,ito2}, 
\be &{\cal D}_1\phi - \omega x^1 B_3 =  \pm B_1&\nn\\ &{\cal D}_2\phi - \omega x^2 B_3 = \pm B_2&  \label{mono}\\ &{\cal D}_3\phi + \omega(x^1 B_1 + x^2 B_2) = \pm B_3&
\nn\ee
where the $\pm$ signs are for monopoles and anti-monopoles respectively. 
Solutions to these equations describe monopoles carrying magnetic charge in $U(1)^{N-1} \subset SU(N)$, trapped at the origin of the $(x^1,x^2)$ plane. These monopoles have mass given by the usual expression
\be M =\frac{1}{e^2} \int d^3x\ \partial_\alpha \, {\rm Tr}\,\phi B_\alpha \label{monomass}\ee
with $\alpha=1,2,3$. 
A study of the simplest, charge one monopole in $SU(2)$ was performed in \cite{ito2}. It was found that the gauge field profile is unchanged by the presence of the trap, while  the scalar profile of the solution is deformed. In particular, in the presence of a monopole, the expectation value of the scalar field $\phi$ differs in different asymptotic parts of space. If we pick an expectation value for $\phi \in su(2)$,
\be \phi \rightarrow  a\,\sigma^3\ \ \ \ {\rm as}\ x^1,x^2 \rightarrow \infty \ {\rm with }\ x^3 =0\nn\ee
then the monopole solution of \cite{ito2} has
\be \phi \rightarrow (a\pm \omega)\, \sigma^3 \ \ \ \ {\rm as}\ x^3\rightarrow \pm\infty\label{phishift}\ee
This doesn't affect the mass given in \eqn{monomass}, which is independent of $\omega$. 

\para
Little appears to be known about the solutions to \eqn{mono} for magnetic charge $n\geq 2$ or, indeed, for higher-rank gauge groups. In the absence of a trap, it is known than there exists a unique axially symmetric $SU(2)$ monopole for each charge $n$ \cite{prasad}.  For $n\geq 2$, the profiles look like \raisebox{-.2ex}{\epsfxsize=0.3in\epsfbox{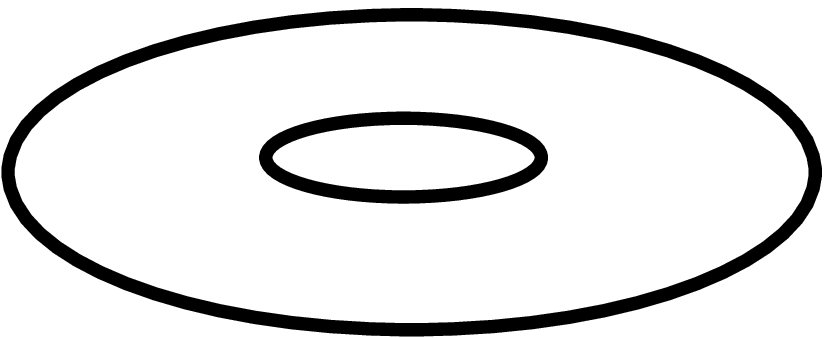}}. It is not clear if the BPS equations \eqn{mono} have a corresponding single solution for charge $n\geq 2$ or whether they admit more general solutions in which $n$ charge one monopoles are free to roam in the $x^3$ direction.

\subsection{Vortices}\label{vsec}

When $v^2\neq 0$, the fundamental scalar fields $q_i$ pick up an expectation value. When $N_f=N$, this is given by
\be \langle q_i^{\,a} \rangle = v \delta_i^{\, a}\label{qvev}\ee
The $U(N)$ gauge symmetry is now  completely broken. In this phase, the theory admits BPS non-Abelian vortex strings, first introduced in  \cite{vib,konishi}. 

\para
In general, the adjoint scalar field $\phi$ also picks up an expectation value
\be \langle \phi \rangle = {\rm diag} (m_1,\ldots, m_N)\label{phivev}\ee
The dynamics of vortices depends on these mass parameters $m_i$. We will start by describing the vortices when $m_i=0$. In Section \ref{confmonsec} we describe vortices with $m_i\neq m_j$.

\para
When $m_i=0$, the theory has an $SU(N)$ global symmetry, in addition to the $U(N)$ gauge symmetry. The expectation value \eqn{qvev} induces the symmetry breaking 
\be U(N)_{\rm gauge} \times SU(N)_{\rm global}\ \longrightarrow\ SU(N)_{\rm diag}\nn\ee
This surviving  $SU(N)_{\rm diag}$ symmetry acts on the vortices, endowing them with internal non-Abelian degrees of freedom.

\para
In the presence of the trap, the vortex strings are only BPS when aligned along the $x^3$ direction. The profile of the fields in the $(x^1,x^2)$ plane obey the vortex equations
\be {\cal D}_z q_i = 0 \ \ \ ,\ \ \ B_3 = e^2 ( q_i q_i^\dagger - v^2)\label{vortex}\ee
where $z=x^1 + i x^2$, together with the supplementary equations for $\phi$, 
\be  {\cal D}_z\phi = \frac{\omega}{2}\bar{z}B_3 \ \ \ ,\ \ \ \phi q_i = \omega \bar{z}\,{\cal D}_{\bar{z}} q_i
\label{vphi}\ee
Solutions to these equations have tension
\be T = -v^2 \int d^2 x\ {\rm Tr}\, B_3 = 2\pi v^2 k\ \ \ \ \ k\in {\bf Z}^+\label{tension}\ee
The situation with vortices is rather more straightforward than the situation with monopoles. The vortex equations \eqn{vortex} remain unchanged which means, in turn, that the profiles for $A_z$ and $q$ remain the same. However, we now must also solve the equations \eqn{vphi} for $\phi$ in the background of this vortex. These have a simple interpretation: they are, respectively, the  requirement that $A_z$ and $q$ are rotationally invariant, up to a gauge transformation. 

\para
The upshot is that, in the presence of the trap, only axially symmetric vortex configurations survive. We flesh out this statement in Appendix \ref{appa}, where we also show that the dynamics of vortices can be described as motion on the original moduli space in the presence of a potential. This potential is generated by the Killing vector associated to planar rotations.

\para
For $U(1)$ vortices, there is a unique axially symmetric configuration for each magnetic flux $k$. The zeros of the Higgs field $q$ all sit at the origin. The magnetic flux is roughly constant with $B\approx - e^2 v^2$ over a disc of radius $R\approx \sqrt{k}/ev$, before dropping exponentially quickly to zero. Solving \eqn{vphi} shows that $\phi$ rises approximately quadratically within this disc, $\phi \approx \omega B |z|^2 /2$, before it too drops exponentially quickly to zero at a radius $|z|\approx R$. 

\para
For vortices in $U(N)$, the situation is more complicated. Each vortex has an internal orientation which, when well separated from other vortices, is parameterised by ${\bf CP}^{N-1}$. However, in the presence of the trap the vortices are pushed on top of each other and these internal orientations mix in a complicated manner. (We will provide a description of this in Section \ref{kinksec}.) For $k=2$ vortices in $U(2)$, the moduli space of axially symmetric vortices was determined in \cite{kojime,moreshif,eto} to be topologically, but not metrically, ${\bf CP}^2/{\bf Z}_2$. Surprisingly, this sub-manifold is singular even though the full moduli space of vortices is smooth.

\subsection{Confined Monopoles}\label{confmonsec}

When $v^2\neq 0$ and $m_i\neq m_j$ the theory admits BPS monopoles, but these are now confined: the magnetic flux leaves the monopoles along the $x^3$ direction as a vortex string. The resulting solitons are vortex/monopole composites. The possibility of such monopoles confined on ${\bf Z}_2$ strings was pointed out in \cite{hind}. Here we will be interested in BPS monopoles confined on non-Abelian strings. These were introduced in \cite{monohiggs} and further explored in \cite{shifyung,meami2,isozumi,auzzi,shifyung2,sakaime,nitta,wimmer0,wimmer}.

\para
To describe these confined monopoles, it is perhaps simplest to first look at what becomes of the vortex strings. With the masses $m_i$ turned on, the symmetry breaking pattern is now
\be U(N)_{\rm gauge}\times SU(N)_{\rm global}\ \stackrel{m}{\longrightarrow}\ U(N)_{\rm gauge}\times U(1)^{N-1}_{\rm global}\ \stackrel{v^2}{\longrightarrow}\ U(1)^{N-1}_{\rm diag}\nn\ee
 where the first of these symmetry breakings is explicit, the second spontaneous. There is no longer a surviving non-Abelian symmetry and, correspondingly, the internal orientation modes of the vortex are lifted. Instead, the vortices sit in the Cartan subalgebra $U(1)^N\subset U(N)$. The magnetic flux $k$ of a vortex is distributed among these $N$ Cartan elements, and the most general solution takes the form 
 \be B_3 = {\rm diag} (B_3^{(k_1)},B_3^{(k_2)},\ldots,B_3^{(k_N)}) \ \ \ {\rm and}\ \ \ q_i^{\, a} = \delta_i^{\, a} q^{(k_a)}\label{abmag}\ee
where $\sum_a k_a = k$ and  $(B_3^{(k)},q^{(k)})$ is the solution for an Abelian vortex with magnetic flux $k$. As we have seen, in the presence of the trap there is a unique Abelian vortex solution with a given magnetic flux.  This means that the non-Abelian vortex ground states are labelled by the ordered set $\{k_a\}$.

\para
The number of ground states for a non-Abelian vortex is the number of ways of distributing the $k$ total fluxes among the $N$ Cartan elements. In other words,
\be \mbox{Number of ground states} = \frac{(k+N-1)!}{k!(N-1)!}\label{combinatorics}\ee
 The confined monopoles are objects which interpolate from one of these ground states to another. They act as sources and sinks for magnetic flux. Because the magnetic charge lies in $U(1)^{N-1} \subset SU(N)$, the monopoles do not change the total flux $k$. They merely redistribute it among the Cartan elements.

\para
In the absence of the harmonic trap, the Bogomolnyi equations for confined monopoles were derived in \cite{monohiggs}. In the presence of a trap, it is straightforward to derive modified Bogomolnyi equations giving solutions to the equations of motion of \eqn{lag}. They are
\be &{\cal D}_1\phi - \omega x^1 B_3 = \pm B_1&\nn\\ &{\cal D}_2\phi - \omega x^2 B_3 = \pm B_2\nn\\ 
& {\cal D}_3\phi + \omega(x^1 B_1 + x^2 B_2) = \pm (B_3 - e^2 (q_iq_i^\dagger - v^2))&\label{confmono}\\ 
&{\cal D}_z q_i = 0&\nn\\ & {\cal D}_3 q_i = \pm ( (\phi-m_i)q_i - \omega \bar{z}{\cal D}_{\bar z}q_i)&\nn\ee
These equations describe BPS vortices (rather than anti-vortices) threaded by either BPS or anti-BPS monopoles, depending on the $\pm$ sign on the right-hand side. Solutions to these equations have string tension \eqn{tension}, together with a finite contribution from the mass of the monopole. We will show that this mass is given in terms of the magnetic fluxes $\{k_a\}$ at $x^3=\pm \infty$ by
\be M_{\rm mono} = \mp \frac{2\pi}{e^2}\sum_a\Bigg[ m_a k_a  - \frac{\omega}{2}k_a^2 \Bigg]^{x^3=+\infty}_{x^3=-\infty}\label{confmonomass}\ee
The first of these terms coincides with the monopole mass \eqn{monomass} when $\phi$ has expectation value \eqn{phivev}, albeit with the flux now collected at $x^3=\pm \infty$ as befits a vortex string, rather than radially for a monopole in the deconfined phase. The second term, quadratic in fluxes, is novel. As we will explain shortly, it describes the binding energy between the monopole and vortex string. 

\subsubsection*{Computing the Mass of the Confined Monopole}

We now derive the formula \eqn{confmonomass}. We look for static solutions with $\partial_0=A_0=0$. We take the energy functional associated to \eqn{lag} and complete the square thus:
%
%
%

%
\be  {\cal E}   =  \int d^3x\ && \frac{1}{2e^{2}}{\rm Tr}\,\left({\cal D}_{1}\phi-\omega x^{1}B_{3}+\epsilon B_{1}\right)^{2}+\frac{1}{2e^{2}}{\rm Tr}\,\left({\cal D}_{2}\phi-\omega x^{2}B_{3}+\epsilon B_{2}\right)^{2} \nn\\
&& +\frac{1}{2e^{2}}{\rm Tr}\,\Big[{\cal D}_{3}\phi+\omega x^{1}B_{1}+\omega x^{2}B_{2}+\epsilon(B_{3}-e^{2}(\sum_{i}q_{i}q_{i}^{\dagger}-v^{2}))\Big]^{2}\nn\\
&& +\sum_{i=1}^N\Bigg\{ |{\cal D}_{z}q_{i}|^{2} + |{\cal D}_{3}q_{i}-\epsilon((\phi-m_{i})q_{i}- i\omega(x^{1}{\cal D}_{2}q_{i}-x^{2}{\cal D}_{1}q_{i}))|^{2}\Bigg\}\nn\\
&& -v^{2}{\rm Tr}\, B_{3} -\epsilon\frac{1}{e^{2}}{\rm Tr}\, B_\alpha{\cal D}_\alpha \phi  -\epsilon v^{2}{\rm Tr}\,{\cal D}_{3}\phi -\epsilon\omega v^{2}{\rm Tr}\,(x^{1}B_{1}+x^{2}B_{2}) \nn\\
&& \nn\\
&& +\sum_{i=1}^N\Bigg\{\epsilon{\cal D}_{3}(q_{i}^{\dagger}(\phi-m_{i})q_{i}) +\epsilon\omega q_{i}^{\dagger}(x^{1}B_{1}+x^{2}B_{2}) q_{i} \nn\\
&& - \epsilon i\omega{\cal D}_{3}q_{i}^{\dagger}(x^{1}{\cal D}_{2}q_{i}-x^{2}{\cal D}_{1}q_{i}) + \epsilon i\omega(x^{1}{\cal D}_{2}q_{i}^{\dagger}-x^{2}{\cal D}_{1}q_{i}^{\dagger}){\cal D}_{3}q_{i}\Bigg\}\nn \ee
where $\epsilon = \pm $ determines whether we're dealing with monopoles or anti-monopoles. The Bogomolnyi equations \eqn{confmono} can be seen in the total squares of the first three lines, while the 
first term of the fourth line is the tension of the vortex string. The remaining terms are the mass of the confined monopole. After integrating by parts, and using ${\cal D}_\alpha B_\alpha=0$, we can write them as 
\be M_{\rm mono} 
&=& \epsilon \int d^3x\  -\frac{1}{e^{2}} {\cal \partial}_{\alpha}{\rm Tr}\,(B_{\alpha} \phi)  -  v^{2}{\rm Tr}\,{\cal D}_{3}\phi - \omega v^{2}{\rm Tr}\,(x^{1}B_{1}+x^{2}B_{2}) \nn\\
&& +\ \sum_{i=1}^N  \partial_{3} \Bigg( q_{i}^{\dagger}(\phi-m_{i})q_{i} -  i\omega q_{i}^{\dagger}(x^{1}{\cal D}_{2} -x^{2}{\cal D}_{1})q_{i} \Bigg)  \label{monopolelagrangian} \ee
All these terms are total derivatives. As we already mentioned, because the flux is collimated in a vortex string, the first term gets contributions from $x^3=\pm \infty$ rather than radially. 
The last of these terms vanishes because, asymptotically, the configuration looks like an axially symmetric vortex string, with $\phi$ obeying \eqn{vphi}. We're left having to evaluate
\be M_{\rm mono} = -\epsilon  \frac{1}{e^2} \int d^2x\ \Big[{\rm Tr}\,\phi( B_3 + e^2 v^2 ) + \omega e^2 v^2 {\rm Tr}\,(-x^{1}A_2+x^{2}A_{1}) \Big]^{x^3=+\infty}_{x^3 = -\infty}\nn\ee
We can massage the last term. We write
\be \omega\int d^2x\ {\rm Tr}\,(-x^1 A_2 + x^2 A_1) = \omega \int d^2x\ \frac{\bar{z}z}{2}{\rm Tr}\,B_3 = \int d^2x\ {\rm Tr}\,\phi \nn\ee
where we've used the expression \eqn{vphi} for the profile of $\phi$ in the background of the vortex. We've also discarded some boundary terms upon integration by parts; these are  common to $x^3=\pm \infty$ and so do not contribute to the monopole mass, which becomes
\be M_{\rm mono} = \mp \frac{1}{e^2} \int d^2x\ \Big[{\rm Tr}\,\phi( B_3 + 2e^2 v^2) \Big]^{x^3=+\infty}_{x^3=-\infty}\label{halfway}\ee
This is pleasingly simple, involving asymptotic integrals at $x^3=\pm \infty$. However, it's not obvious that it depends only on topological information. Indeed, this is even true of the first term which, naively, looks the same as the usual monopole mass \eqn{monomass}. The reason for this is that, as we saw in Section \ref{vsec}, the field $\phi$ gets a profile in the presence of a vortex. This profile is complicated, determined by the equations \eqn{vphi}, and feeds into the expression for the mass. Nonetheless, as we now show, nice things happen. 

\para
The nice things follow from the fact that the asymptotic vortices are axially symmetric, so that both $|q|^2$ and $\phi$ are a function only of the radial coordinate $r$. Consequently, it is straightforward to show
\be \partial_r \phi = \omega r B_3\nn \ee
with $\phi \rightarrow \langle\phi\rangle = {\rm diag}(m_1,\ldots,m_N)$ as $r\rightarrow \infty$, and $\phi\rightarrow \langle \phi\rangle + \omega\,{\rm diag}(k_1,\ldots, k_N)$ as $r\rightarrow 0$, a result which was also used in \cite{hengyu}. Using all these facts, we have
\be \int d^2 x \ {\rm Tr}\, \phi B_3 =  2 \pi \ \Big[ \frac 1 {2\omega}  \phi^2 \Big]_{r=0}^{r=\infty} = -2\pi \sum_i \Big(k_i m_i +\frac{\omega}{2} k _i^2 \Big) \label{bphi} \ee
We're left with having to compute the linear term, $\int d^2x\ {\rm Tr}\,\phi$ in \eqn{halfway}. This follows straightforwardly from the vortex equations \eqn{vortex} and \eqn{vphi}, which imply

\para
\be  {\rm Tr}\, \phi( B_3 + e^2 v^2 )& =& e^2 \sum_i q_i^\dagger \phi q_i \nn\\ &=& e^2 \sum_i   m_i q_i^\dagger q_i + \omega q_i^\dagger \bar{z} {\cal D}_{\bar{z}} q_i   \nn\\ 
&=& \sum_i m_i \left(  B_3^{(i)} + e^2 v^2\right) + \omega e^2{\cal D}_{\bar{z}}( \bar{z}(q_i^\dagger q_i - v^2)) - \omega e^2 (q_i^\dagger q_i - v^2)\nn\\
&=& \sum_i m_i\left( B_3^{(i)} + e^2v^2\right) - {\omega} B_3^{(i)} \nn \ee
where the magnetic field has been decomposed into its Abelian components \eqn{abmag}. The integrals can now be expressed in terms of the Abelian fluxes $\{k_a\}$. Dropping terms  common to the two boundaries at $x^3=\pm \infty$, we have
%
%
%
\be  \int d^2x\ \Big[{\rm Tr}\,\phi( B_3 + e^2 v^2 ) \Big]^{x^3=+\infty}_{x^3 = -\infty} = -2\pi \sum_i \Big[ m_i k_i \Big]^{x^3=+\infty}_{x^3 = -\infty}  \nn \ee
%
%
%
Comparing to \eqn{bphi}, we learn that
\be e^2 v^2 \int d^2x\ \Big[ {\rm Tr}\,\phi\Big]^{x^3=+\infty}_{x^3=-\infty} = +2\pi \sum_i \Big[\frac{\omega}{2}k_i^2\Big]^{x^3=+\infty}_{x^3=-\infty} \nn\ee
Finally, we can put this all together to give 
\be M_{\rm mono} = \mp \frac{2\pi}{e^2}\sum_i\Bigg[ m_i k_i  - \frac{\omega}{2}k_i^2 \Bigg]^{x^3=+\infty}_{x^3=-\infty}\label{massagain}\ee
which is the promised result \eqn{confmonomass}. 

\subsubsection*{Interpreting the Monopole Mass}

The term linear in $k$ is the familiar monopole mass term. However, it is unusual to find BPS solitons whose mass is quadratic in the charges.
Here we provide an interpretation of the quadratic term in \eqn{massagain}. We will argue that it can be thought of as a binding energy between the confined monopole and vortex strings.

\para
To see this, it will suffice to work with $U(2)$ gauge theory. We choose the masses to be $m_1 = -m$ and $m_2= +m$, so that the asymptotic expectation value is $\langle \phi\rangle = -m \sigma^3$. Let's first look at the charge $k$ monopole, interpolating between the vortex strings with flux $(k_1,k_2) = (0,k)$  at $x^3=-\infty$ and flux $(k,0)$ at $x^3=+\infty$. The confined monopole looks like:
\be \raisebox{1.1ex}{\epsfxsize=2.7in\epsfbox{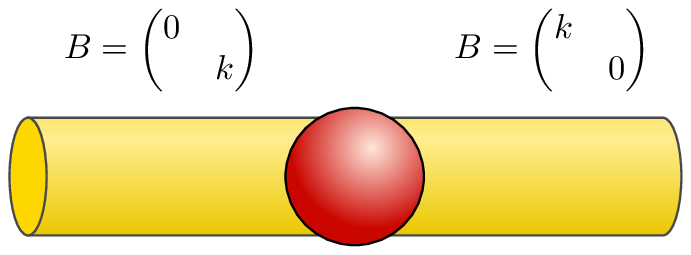}}
 \nn\ee
The mass of the monopole is given by
\be M_{\rm mono} = \frac{4\pi m k}{e^2}\label{easymass}\ee
This is the usual mass of a monopole in an $SU(2)$ gauge theory. There is no contribution from the harmonic trap.

\para
Let's now look at the same monopole of charge $k$, this time interpolating between $(p_1,k+p_2)$ at $x^3=-\infty$ and $(k+p_1,p_2)$ at $x^3=+\infty$. We can think of this as a combination of the previous configuration, superposed with a vortex with flux $(p_1,p_2)$.
\be \raisebox{-3.1ex}{\epsfxsize=5.5in\epsfbox{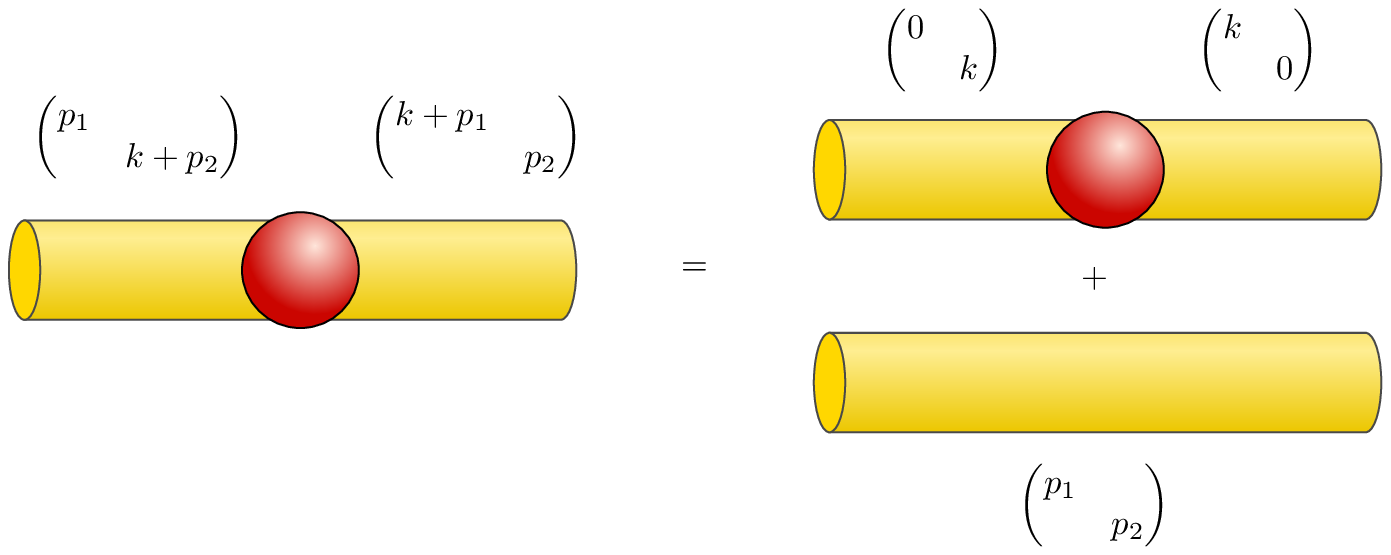}} \nn\ee
The mass of the monopole is now
\be 
M_{\rm mono} = \frac{4\pi mk}{e^2} + \frac{2\pi \omega k}{e^2} (p_1-p_2)\nn\ee
The second term is due to the presence of the $(p_1,p_2)$ vortex string and, in this sense, can be thought of as the binding energy between the monopole and this string. Note that this binding energy can be positive or negative. Note also that $\omega$ can take either sign. 

\para
There are further questions that we would like to ask. Let's return to the charge $k$ monopole with mass \eqn{easymass}. Can this monopole decompose into charge one constituents? The first few monopoles in the string would look like this:
\be \ \nn \raisebox{-5.1ex}{\epsfxsize=4in\epsfbox{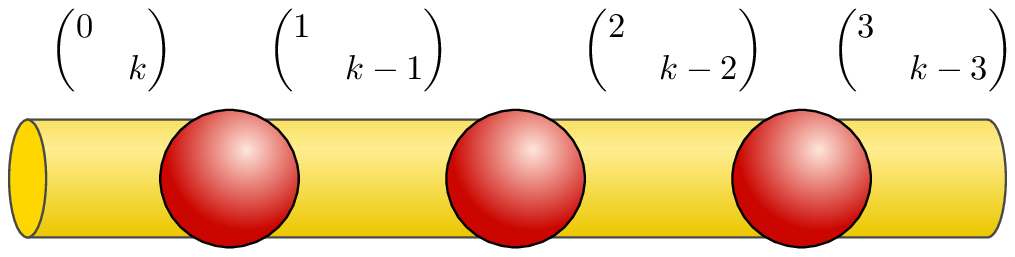}}\nn \\
 \nn\ee
If this is possible, each charge one monopole now experiences a binding energy with the underlying vortex string. The mass of the $q^{\rm th}$ monopole in the sequence ($q=1,\ldots, k)$ is given by
\be M_q = \frac{4\pi m}{e^2} - \frac{2\pi \omega}{e^2} (k-2q+1)\label{massq}\ee
Although the mass of each individual constituent gets a contribution from the binding energy, the sum of all these masses is again \eqn{easymass}, independent of $\omega$. 

\para
There's something unsettling about the mass formula \eqn{massq}. 
For the sake of the argument, let's choose $\omega>0$.  The first monopoles in the sequence \eqn{massq} then have negative binding energy; the later ones positive. More worryingly, if we take this formula at face value then it  looks as if the mass of the first monopoles will become negative for sufficiently large $k$.  This is clearly unacceptable. 

\para
In the next section we will understand what's going on with this necklace of monopoles. We will show that a charge $k$ monopole can, in fact, decompose into its constituent parts as described above, but this can only happen if the masses of those constituents are positive.  Moreover, the story for higher rank gauge groups will turn out to be somewhat more involved.

\section{Monopoles as Kinks on the Worldsheet}\label{kinksec}

In this section, we explore the confined monopoles in more detail. However, instead of solving the Bogomolnyi equations \eqn{confmono}, we will instead look at the monopoles from the perspective of the vortex string worldsheet, where they appear as BPS kinks. This was first shown in \cite{monohiggs}, and further explored in \cite{shifyung, meami2,hananytong}. However, attention has previously focussed on monopoles threaded on a single $k=1$ vortex string. These are restricted to a single monopole in $U(2)$ gauge theories or the so-called $\vec{g} = (1,1,\ldots,1)$ monopoles in $U(N)$ theories. 

\para
Here we will be interested in monopoles threaded on more than one vortex string. In part, the problem with studying this in the past was to decouple the effect of the string positions from the dynamics of the monopoles along in the string. Here, the harmonic trap helps us since, as we have seen, the strings are obliged to sit at the origin.

\subsection{A Matrix Model for Vortex Strings}

We could study the vortex string dynamics from the geometric perspective of the moduli space, as described in Appendix \ref{appa}. However, a more versatile description is offered by the vortex matrix model introduced in \cite{vib}. 

\para
The matrix model provides a description of the vortex dynamics in terms of an ${\cal N}=(2,2)$ gauged linear sigma model. It is a $U(k)$ gauge theory, with a single complex adjoint scalar $Z$ and $N$ fundamental scalars $\varphi_i$, $i=1,\ldots, N$. We also include an auxiliary adjoint scalar, $\sigma$.  Neglecting fermions, the low-energy vortex dynamics is then described by
\be S_{\rm vortex} = \int d^2x\ && {\rm Tr} \,|{\cal D}_\alpha Z|^2 + \sum_{i=1}^N|{\cal D}_\alpha \varphi_i|^2 - \frac{g^2}{2}{\rm Tr}\,([Z,Z^\dagger] + \sum_i\varphi_i\varphi_i^\dagger - r)^2 \nn\\ && +\,\frac{1}{2g^2}{\rm Tr}\,({\cal D}_\alpha \sigma)^2 -{\rm Tr}\,|[\sigma,Z]+\omega Z|^2 - \sum_{i=1}^N \varphi_i^\dagger(\sigma-m_i)^2\varphi_i \ \ \ \ \label{vaction}\ee
Here $\alpha=0,3$ runs over the vortex string worldsheet directions. Note that although $\sigma$ is generally a complex adjoint scalar, in the present case where we have restricted to real masses $m_i$, we can take  $\sigma$ to be real.

\para
We should think of this action in the $g^2\rightarrow \infty$ limit. (There is also a kinetic term for the $U(k)$ gauge field which we have omitted; it is suppressed in this limit.) The ``D-term" potential on the first line is then imposed  as a $k\times k$ matrix constraint
\be [Z,Z^\dagger] + \sum_{i=1}^N \varphi_i\varphi_i^\dagger = r {\bf 1}_k\label{vconstraint}\ee
If we further quotient by the $U(k)$ gauge action, $Z\rightarrow U Z U^\dagger$ and $\varphi\rightarrow U\varphi$, this defines the vortex moduli space ${\cal M}_{N,k}$. It has dimension
\be {\rm dim}({\cal M}_{N,k}) = 2Nk\nn\ee
Note that this quotient construction defines a metric on ${\cal M}_{N,k}$. This does not coincide with the metric on the vortex moduli space defined in Appendix \ref{appa}. Nonetheless, this construction will suffice for our purposes. In particular, the masses of BPS states coincide with those in the vortex theory if we fix the K\"ahler class of ${\cal M}_{N,k}$ to be \cite{vib}
\be r = \frac{2\pi}{e^2}\nn\ee
The  auxiliary scalar $\sigma$ can be ignored if $\omega=m_i=0$. In this case, after the quotient above, the action \eqn{vaction} becomes a non-linear sigma model on ${\cal M}_{N,k}$. However, if either $\omega\neq 0$ or $m_i\neq 0$ then $\sigma$ is non-vanishing and the last two terms in \eqn{vaction} can be thought of as inducing a potential on ${\cal M}_{N,k}$.

\subsubsection*{U(1) Vortices}

Let us look at vortices in the $U(1)$ theory in more detail. Here there is a single $\varphi$ field in the action \eqn{vaction}.  We may set the mass of $\varphi$ to zero through a  constant shift of $\sigma$. The ground state of $k$ vortices is then given by the constraint \eqn{vconstraint}, supplemented by
\be [\sigma,Z] + \omega Z = 0\label{wzero}\ee
There is a unique solution to these two equations, found in \cite{alexios}.  This is
\be Z_k = \sqrt{r}\left(\begin{array}{cccccc}  0\ & 1\ &  & &  \\  & 0\ & \sqrt{2} & &  \\   & & & \ddots &   \\ &  &  & 0 & \sqrt{k-1} &  \\  & & & & 0  \end{array}\right)\ \ \ {\rm and}\ \ \ \varphi = \sqrt{r} \left(\begin{array}{c} 0 \\  0 \\ \vdots \\ 0\\ \sqrt{k}\end{array}\right)\label{gs}\ee
with $\sigma = \sigma_k \equiv -\omega\,{\rm diag} (k-1, k-2,\ldots, 2,1,0)$. This corresponds to the unique, rotationally invariant $k$-vortex solution.

\subsubsection*{U(N) Vortices}

For $U(N)$ vortices, the situation depends on the masses $m_i$. For $m_i=0$ (or, indeed, $m_i$ all equal) there remains a moduli space of solutions to \eqn{vconstraint} and \eqn{wzero}.

\para
Here, we are instead interested in the case of distinct masses $m_i\neq m_j$. There are now isolated solutions to \eqn{vconstraint} and \eqn{wzero} in which the total flux $k$ is distributed among the $N$ Cartan elements, mimicking our discussion in Section \ref{confmonsec}. The general solution is determined by an ordered set $\{k_i\}$ with $\sum_i k_i = k$. The matrix $Z$ is block diagonal, while $\sigma$ is diagonal, although conveniently written in block diagonal form
\be Z = \left(\begin{array}{cccccc}  Z_{k_1}\ &  &  &   \\   & Z_{k_2} & &  \\    & & \ddots &   \\ &    &  & Z_{k_N}   \end{array}\right)\  \ \ {\rm and}\ \ \ 
\sigma= \left(\begin{array}{cccccc} \sigma_{k_1} + m_1\ &  &  &   \\   & \sigma_{k_2} + m_2 & &  \\    & & \ddots &   \\ &    &  & \sigma_{k_N} + m_N  \end{array}\right)\ \label{vacua}\ee
with $Z_k$ given by \eqn{gs} and $\sigma_k$ by the expression below \eqn{gs}. Similarly, each $\varphi_i$ has only one non-zero element,  $\sqrt{rk_i}$. 
The number of solutions of this form is given by \eqn{combinatorics}.

\subsection{Kinks on the Worldsheet}

When $\omega\neq 0$ and $m_i\neq m_j$, the worldsheet theory has a large number of isolated ground states. There exist BPS kinks which interpolate between these some of these ground states. These obey the Bogomolnyi equations
\be &{\cal D}_3\varphi_i= (\sigma-m_i)\varphi_i& \nn\\ 
&{\cal D}_3Z = [\sigma,Z] +\omega Z &\label{kink}\\ & {\cal D}_3 \sigma = g^2 ([Z,Z^\dagger] + \sum_i\varphi_i\varphi_i^\dagger -r ) &\nn\ee
BPS domain wall equations with a similar structure have been studied previously, both in Abelian \cite{kinky,medomainwall} and in non-Abelian \cite{nittawall} gauge theories. The novelty here is the presence of the adjoint scalar $Z$. As we will see, this changes the dynamics of these kinks.

\para
Writing the energy as a sum of squares, it is simple to show that the mass of any kink obeying these equations  is given by
\be M_{\rm kink} = -\Bigg[\sum_i m_i \varphi^\dagger_i\varphi_i -\omega {\rm Tr}\, Z^\dagger Z \Bigg]^{x^3=+\infty}_{x^3=-\infty}\nn\ee
We can evaluate this on the ground states \eqn{vacua}. In the vacuum $\{k_i\}$, we have  $\varphi_i^\dagger \varphi_i = rk_i$,  and
\be {\rm Tr} Z^\dagger Z = \frac{r}{2}\sum_{i=1}^N k_i(k_i-1) = \frac{r}{2}(\sum_i k_i^2 - k)\nn\ee
The mass of the kink is therefore given by
\be M_{\rm kink} = -r \sum_{i=1}^N \Bigg[  m_i  k_i - \frac{\omega}{2} k_i^2 \Bigg]^{x^3=+\infty}_{x^3=-\infty} \label{kinkmass}\ee
Identifying $r=2\pi/e^2$, this agrees with the mass of the monopole \eqn{monomass}.

\subsection{Counting Zero Modes for Confined Monopoles}

We now return to the question that we asked in Section \ref{confmonsec}: can a monopole of given charge decompose into constituent monopoles, spread like a necklace of beads along a vortex string? To answer this question, we attempt to count the number of zero modes of the kinks. We do this by computing the index for the linearised domain wall equations \eqn{kink}. This calculation is presented in Appendix \ref{indexapp}. The result is:
\be {\cal I} = \frac{1}{2} \left[ \sum_{i,j}
\sum_{q=0}^{k_i-1}
{\rm sign}(m_j - m_i + (q-k_j)
\omega)\right]_{x^3=-\infty}^{x^3=+\infty}\label{index}
\ee
The sum is over all $i,j$ (i.e. each pair is counted twice). 
It is  understood that there is no contribution from any sector where $k_i=0$.

\para
We will now illustrate the physics underling this index theorem by giving a few simple examples.  But first we give some important warnings. The index counts the number of zero modes of the operator $\Delta$ arising from linearised Bogomolnyi equations, minus the number of zero modes of the adjoint operator $\Delta^\dagger$. (Both of these operators are defined in Appendix \ref{indexapp}.) Usually, when computing the index for BPS solitons,  one can show that the adjoint operator has no zero modes. This means that the index counts what we care about:  the number of collective coordinates of the soliton. This is the case, for example, in \cite{erick,ejw1,ejw2,lee} and \cite{sakaime}. It is not, however, the case here. There are regimes of the parameters $m_i$ and $\omega$ for which we can show that the adjoint operator {\it does} have zero modes. This means that the index ${\cal I}$ can only give a lower bound on the number of collective coordinates. 

\para
On top of this, although we can count the number of zero modes of solutions, we have no proof that solutions actually exist, at least  beyond the simplest cases with $Z=0$ \cite{sakaiyang}. These issues mean that any attempt to extract physics from \eqn{index} alone will be far from rigorous. Nonetheless, if we assume that solutions to \eqn{kink} exist whenever ${\cal I}>0$, then a consistent story emerges from a combination of the index and the mass formula \eqn{kinkmass}. We now tell this story.

%
%
%

\subsubsection*{Confined Monopoles in $U(2)$}

We start by revisiting the problem that we introduced in Section \ref{confmonsec}: is it possible for a charge $k$ $SU(2)$ monopole to decompose into its constituent charge one components? The first few monopoles of the resulting configuration would look like this:
\be \raisebox{-5.1ex}{\epsfxsize=4in\epsfbox{flux3.eps}}\nn\\
 \nn\ee
Thankfully, for $U(2)$ monopoles, or domain walls with two flavours, the adjoint operator $\Delta^\dagger$ has no zero modes, so we take the index ${\cal I}$ at face value as counting the number of collective coordinates of our confined monopole.

\para
As in Section \ref{confmonsec}, we set $m_i=(-m,+m)$ and take $\omega >0$. The index theorem \eqn{index} becomes
\be {\cal I} = \frac{1}{2}\sum_{q=0}^{k-1}\left[{\rm sign}(2m+q\omega) - {\rm sign}(-2m + q\omega)\right]  = 
\frac{k}{2} - \frac{1}{2}\sum_{q=0}^{k-1}\left[{\rm sign}(-2m+q\omega)\right] \nn\ee
We see that the number of zero modes depends on the relative values of $m$ and $\omega$. In what follows, we fix the masses and see how the index varies as we change $\omega$. Suppose that we start with small $\omega$, so the trap is weak. 
Then we have
\be (k-1)\omega < 2m \ \ \ \Rightarrow\ \ \  {\cal I} = k\nn\ee
This counts complex zero modes. It strongly suggests that the charge $k$ monopole can indeed split into its constituents. Each of these constituents has a position, and a phase. The phase associated to domain walls was first discussed in \cite{pauled} and is entirely analogous to the phases associated to monopoles; in both cases, excitations of this phase coordinate turn the monopoles into dyons.

\para
However, as we increase the strength of the trap, interesting things happen. We see from \eqn{massq} that at the threshold $(k-1)\omega=2m$ the mass of the first monopole vanishes: $M_1=0$. Moreover, the index theorem \eqn{index} tell us that, at this point, the number of zero modes jumps by one:
\be (k-2)\omega < 2m < (k-1)\omega\ \ \ \Rightarrow\ \ \ {\cal I} = k-1\nn\ee
What's happening here is the following: as $(k-1)\omega\rightarrow 2m$, the left-hand monopole becomes massless and swells to infinite size, joining with the second monopole on the string. The first few monopoles are now
\be \raisebox{-2.1ex}{\epsfxsize=4in\epsfbox{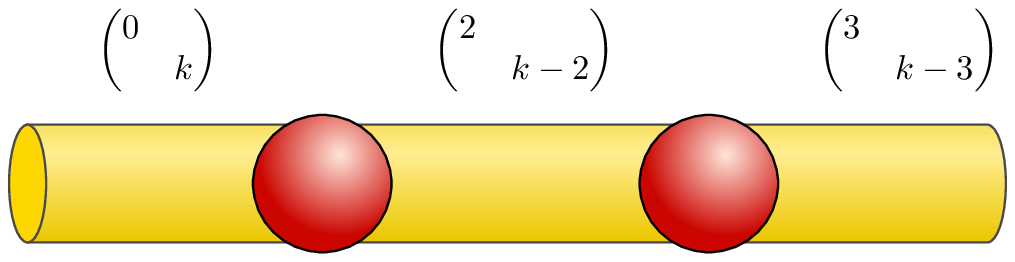}} \nn\\
 \nn\ee
The first monopole on the string now has charge 2, and the others charge 1. From \eqn{confmonomass}, the mass of this charge 2 monopole is
\be M'= \frac{4\pi}{e^2}\left(2m -\omega(k-2)\right)\nn\ee
The masses of the subsequent monopoles are still given by \eqn{massq}. 
As we lower the mass still further, we again hit a threshold where this charge 2 monopole becomes massless and the index \eqn{index} jumps once again
\be (k-3)\omega <  2m  < (k-2)\omega\ \ \ \Rightarrow\ \ \ {\cal I} = k-2\nn\ee
This pattern now continues. The necklace consists of a charge 3 monopole followed by a string of charge 1 monopoles. As the mass decreases, it is always the left-most, higher charge monopole which becomes massless. In general, if the mass lies in the range
\be (k-p-1)\omega < 2m < (k-p)\omega\ \ \ \Rightarrow\ \ \ {\cal I} = k-p\nn\ee
there are $k-p$ constituent objects, which arrange themselves on the string as a single charge $p+1$ object, followed by 
followed by $k-p-1$ charge one monopoles.  By the time we get to $\omega > 2m$, there is just a single zero mode left.

\para
We learn that a monopole can split into constituent parts only if the mass of these constituents is positive. As we increase the strength of the harmonic trap, the quadratic binding energy causes more and more monopoles to become massless and there are fewer physical constituents. 
It seems plausible that the jump of the index at integer values of $2m/\omega$ is related to the shift of the asymptotic value of $\phi$ \eqn{phishift} found in \cite{ito2}. It would be interesting to understand this connection better.

\subsubsection*{Confined Monopoles in $U(3)$}

We now turn to monopoles in $U(3)$. Here, for the first time, we run into the issue of zero modes for the adjoint operator $\Delta^\dagger$, which means that the index \eqn{index} undercounts the number of zero modes. For this reason, we go slowly and present three examples of successive complexity. In all of these, we again order the masses such that $m_{i+1}>m_i$.

\para

\ \\ 
\underline{Example 1:}

\para
We start by considering the example of flux $\{k_i\} = (0,1,1)$ at $x^3\rightarrow -\infty$ interpolating  to $\{k_i\} = (1,0,1)$ at $x^3=+\infty$. We write this as 
\be (0,1,1)\ \  \longrightarrow \ \ (1,0,1)\nn\ee
This is simply a $U(2)$ monopole embedded in the upper-left  corner of $U(3)$, now accompanied by a single flux tube in the lower-right corner. One might think that these two objects don't talk to each other and, indeed, for most values of $\omega$ the index is ${\cal I}=1$. But, for $\omega$ in a particular range, the index tells us that something interesting can happen. We have
\be {\cal I} = 1 + \frac{1}{2}\left[{\rm sign}(m_3-m_1-\omega) - {\rm sign} (m_3-m_2-\omega)\right] =  \left\{\begin{array}{ll} 1\ \ \ \ \ \ \ \ \  &\omega  < m_3-m_2 \nn\\ 
1\ \ \ &\omega > m_3-m_1\nn\\ 2 & {\rm otherwise}\nn\end{array}\right.
\nn\ee
In  the window $(m_3-m_2)<\omega<(m_3-m_1)$, it appears that this single monopole can decompose into two constituents. We believe that this is indeed the correct interpretation. A study of the asymptotics of fluctuations around \eqn{kink} suggest that, for this range of $\omega$, the scalar $Z$ is not just a bystander in the kink equations; its off-diagonal components can turn on. This corresponds to a the charge one monopole decomposing into the following two constituents:
\be (0,1,1)\ \  \longrightarrow \ \  (0,0,2) \ \  \longrightarrow\ \ (1,0,1)\nn\ee
At first glance, this looks unlikely. The first monopole has charge $(0,-1,1)$, the second charge $(1,0,-1)$.  But naively (where ``naive" means ``in the absence of the trap"), the second monopole would be BPS and the first anti-BPS. It is very unusual for these two monopoles to sit happily next to each other because they carry opposite magnetic charge in the final component.  (Something similar, although not entirely the same, happens with calorons \cite{caloron1,caloron2}.)

\para
To see why this is acceptable in the harmonic trap, let us look at the mass formula \eqn{kinkmass}. This is the mass formula for a BPS monopole. Usually,  anti-monopoles would have negative mass when evaluated on this formula, reflecting the fact that we should use the anti-BPS mass formula which differs by a minus sign. However, we find that the mass of the monopole interpolating from $(0,1,1)\rightarrow (0,0,2)$ is indeed positive when $\omega > (m_3-m_2)$.

\para
The emergence of this new monopole of positive mass explains the jump in the index. But why does it subsequently decrease back to ${\cal I}=1$ when $\omega > (m_3-m_1)$? This can be seen by looking at the second monopole, interpolating from $(0,0,2)\rightarrow (1,0,1)$. From \eqn{kinkmass}, we learn that this monopole has negative mass when $\omega > (m_3-m_1)$. In this regime, we are back down to a single bead on the vortex string.

\ \\ \ 
\underline{Example 2:}

\para
Let us now consider a very similar example, 
\be (1,0,1)\ \  \longrightarrow \ \ (1,1,0)\nn\ee
This is again a $U(2)$ monopole accompanied by a $U(1)$ flux in a different sector. As in our previous example, there is a window of $\omega$ in which the index exhibits a jump. Except, this time it goes the other way:

\be {\cal I} = 1 + \frac{1}{2}\left[{\rm sign}(m_2-m_1-\omega) - {\rm sign} (m_3-m_1-\omega)\right] =  \left\{\begin{array}{ll} 1\ \ \ \ \ \ \ \ \  &\omega  < m_2-m_1 \nn\\ 
1\ \ \ &\omega > m_3-m_1\nn\\ 0 & {\rm otherwise}\nn\end{array}\right.
\nn\ee
In the window $(m_2-m_1)<\omega<(m_3-m_1)$, the index vanishes. However, we know that this cannot be a correct count of the number of zero modes because it is always possible to embed the known $U(2)$ solution, with its single complex zero mode, in the lower-right corner of the $U(3)$ matrix. 

\para
What's happening here is that, in this window of $\omega$, the adjoint operator $\Delta^\dagger$ gains a zero mode. (Again, this conclusion is reached by investigating the asymptotics of the appropriate linearised equations.) This is not a physical fluctuation, but is subtracted from the index. This  means that the index is under counting. For this example, the correct number of (complex) collective coordinates of the confined monopole is 1 for all values of $\omega$. 

\ \\ \ 
\underline{Example 3:}

\para
As our final example, we consider a $U(3)$ monopole which, in the absence of a trap, would decompose into constituents \cite{gno,ejw2}. This is the monopole
\be (0,1,1) \ \ \longrightarrow\ \ (1,1,0)\nn\ee
The index \eqn{index} is
\be {\cal I} = 2 + \frac{1}{2}\left[{\rm sign}(m_2-m_1-\omega) - {\rm sign} (m_3-m_2-\omega)\right] \nn\ee
For small $\omega$, the index is ${\cal I}=2$ suggesting that the monopole can split into two constituents. These are naturally identified as the $(1,-1,0)$ monopole and the $(0,1,-1)$ monopole. There are two possible orderings of these constituents on the vortex string, resulting in the sequence of flux tubes
\be  (0,1,1) \ \ \longrightarrow\ \ (0,2,0) \ \ \longrightarrow\ \ (1,1,0)\nn\\   (0,1,1)\ \ \longrightarrow\ \ (1,0,1)\ \ \longrightarrow\ \ (1,1,0)\label{ordering1}\ee
We conjecture that both orderings are allowed. In other words, there is a moduli space of monopoles of  these two different types and these two monopoles can pass through each other.

\para
Now we increase $\omega$. 
There is a window where the index either jumps up to 3 or down to 1. When the index rises, we would like to understand physically what this corresponds to. When the index drops, we would like to understand if this reflects a genuine drop in the number of collective coordinates or if it is instead a fake drop caused by the emergence of zero modes for the adjoint operator. 

\para
The physics is different depending on the specific values of the masses. There are two cases to consider:

\ \\ \ 
\underline{Case A: $m_2-m_1< m_3-m_2$}

\para
In the window, $(m_2-m_1)<\omega<(m_3-m_2)$, the index decreases to ${\cal I}=1$. Following our discussion of $U(2)$ monopoles, we'd like to understand this in terms of one of the constituents becoming massless.  Indeed, if we look at the first route in the ordering \eqn{ordering1}, we see that the monopole $(0,2,0)\rightarrow (1,1,0)$ does indeed become massless at $\omega = (m_2-m_1)$ as we enter this window. This means that, inside the window, we cannot decompose the monopoles in the first ordering. 

\para
However, none of the monopoles in the second ordering in \eqn{ordering1} become massless as we enter the window. From this, we conjecture that the decrease in the index is actually a fake: the number of collective coordinates does not decrease at $\omega=(m_2-m_1)$. Instead, the decrease in the index is telling us that only the second ordering of monopoles is allowed.  In fact, the two constituent monopoles in the second route remain massive for all values of $\omega$. We conjecture that the constituent structure of monopoles in this case is given by
\be \begin{array}{cl} \omega < (m_2-m_1) \ \ \ \ &\mbox{Two constituents, with either of the orderings \eqn{ordering1}}
\\
\omega > (m_2-m_1)   \ \ \ \ &\mbox{Two constituents, with only the second ordering in \eqn{ordering1}}
\end{array}\nn\ee

\ \\ \ 
\underline{Case B: $m_3-m_2< m_2-m_1$}

\para
In the window, $(m_3-m_2)<\omega<(m_2-m_1)$, the index now increases to ${\cal I}=3$. As in previous examples, the lower edge of this window reflects the emergence of a new monopole with positive mass. There are now three constituents, but there remain only two possible orderings because the newly emerged monopole only has positive mass if it sits first on the string:  
\be  (0,1,1) \ \ \longrightarrow \ \ (0,0,2)\ \ \longrightarrow\ \ (0,2,0) \ \ \longrightarrow\ \ (1,1,0)\nn\\   (0,1,1)\ \ \longrightarrow\ \ (0,0,2)\ \ \longrightarrow\ \ (1,0,1)\ \ \longrightarrow\ \ (1,1,0)\label{ordering2}\ee
The first monopole in the sequence has charge $(0,-1,+1)$ and has positive mass for $\omega > (m_3-m_2)$. 

\para
We now increase $\omega$ further to reach the upper end of the window at $\omega = (m_2-m_1)$. If we have taken the first route, the final monopole $(0,2,0)\rightarrow (1,1,0)$ becomes massless at this point and must merge with the middle monopole, leaving us with the sequence $ (0,1,1)\rightarrow (0,0,2)\rightarrow (1,1,0)$. However, this is a subsequence of the second route. And, from the perspective of the second route, nothing bad happens at the point $\omega = (m_2-m_1)$. We therefore suspect that this upper end of the window is a fake and the number of zero modes does not decrease from 3 back to 2 at $\omega = (m_2-m_1)$.

\para
However, this is not to say that the number of zero modes remains 3 for all $\omega$. Something interesting does happen at the later point $\omega = (m_3-m_1)$. Here the monopole $(0,0,2)\rightarrow (1,0,1)$ becomes massless. 

\para
Putting these facts together leaves us with the following conjecture for the decomposition of monopoles in this system 
\be \begin{array}{cl} \omega < (m_3-m_2) \ \ \ \ &\mbox{Two constituents, with either \eqn{ordering1} ordering}
\\
(m_3-m_2) < \omega < (m_2-m_1) \ \ \ \ &\mbox{Three constituents, with either \eqn{ordering2} ordering}
\\
(m_2-m_1) < \omega < (m_3-m_1) \ \ \ \ &\mbox{Three constituents, with only the second \eqn{ordering2} ordering}\\
\omega > (m_3-m_1) \ \ \ \ &\mbox{Two constituents, with only the second \eqn{ordering1} ordering}
\end{array}\nn\ee
Notice that, for both Case A and Case B, we have found ourselves in the same place for large $\omega$, with a restriction on the ordering of the monopoles. 

\subsubsection*{The General Story}

The examples described above seem to capture the kinds of behaviour that can arise for monopoles in $U(N)$ gauge groups.  In general, we conjecture the following rules
\begin{itemize} 
\item If the index increases, it can be traced to a new constituent monopole moving from negative to positive mass.
\item If, for all orderings, some constituent monopole becomes massless, then the number of collective coordinates decreases.
\item If the index decreases and a constituent monopole becomes massless only for some orderings, then those orderings are no longer allowed. The number of collective coordinates does not decrease.
\end{itemize}

\section{Summary and Discussion}

We hope that the discussion above has illustrated the rich dynamics of confined monopoles in the presence of a harmonic trap. Clearly there is much still to understand. We have formulated a conjecture for how monopoles decompose into different constituents with different orderings but, because the index is not equal to the number of collective coordinates, we have had to rely on some guesswork and physical intuition. Ideally, this guesswork can be placed on firmer footing. 

\para
If our conjectures are correct, then the next question is to get a better handle on the moduli space of these confined monopoles and, in particular, how this moduli space changes as we vary $\omega$. 

\para
For a small subset of confined monopoles,  the moduli space has been well studied \cite{nittawall,sakaime}. These are the kinks with $Z=0$. They only exist when the number of vortices is no greater than the rank of the gauge group, $k\leq N$, and even then, describe only a small number of the possible confined monopoles. 
It was shown in \cite{hananytong}, there is a close connection between these $Z=0$ kinks and magnetic monopoles in the absence of a trap. The kink moduli space was shown to be a middle-dimensional complex submanifold of the corresponding monopole moduli space. This submanifold is defined as the fixed point of a $U(1)$ isometry which rotates the monopole configurations in the $(x^1,x^2)$ plane.

\para
One may wonder whether there is a similar connection between free monopoles and the more general confined monopoles with $Z\neq 0$. If such a connection exists, it must be somewhat more complicated. It is known, for example, that the Atiyah-Hitchin moduli space, describing charge two $SU(2)$ monopoles, has a unique fixed point under rotations. Yet we have conjectured above that the corresponding confined monopoles have a moduli space in which the constituents can move independently along the string.  To find a connection in the general case, one should presumably first get a  better handle on the space of solutions to the monopole equations \eqn{mono} in the presence of trap.

\section*{Acknowledgements}

We would like to thank  Matthew Bullimore, Nick Dorey,  Kimyeong Lee, and Nick Manton useful comments and discussions and Alex Considine for grammatical assistance. We're grateful to KIAS for hospitality during the completion of this work. We are supported by STFC and by the European Research Council under the European Union's Seventh Framework Programme (FP7/2007-2013), ERC grant agreement STG 279943, ``Strongly Coupled Systems".

\appendix

\section{Appendix: Dynamics on the Vortex Moduli Space}\label{appa}

In this appendix we show that the dynamics of vortices in the presence of the $\Omega$-deformation is governed by motion on the usual vortex moduli space, but with the addition of a potential. This potential is the length-squared of a Killing vector, $K$, associated to action of rotations.

\para
This result is not surprising and no doubt known to experts. Indeed, the original purpose of the $\Omega$-deformation was to introduce exactly such a potential on the instanton moduli space \cite{nikitaisitcold}. Nonetheless, to our knowledge an explicit derivation has not appeared previously in the literature.  Here we closely follows the method of \cite{potential} which showed how related potentials on other soliton moduli spaces can arise. 

\para
Let us start by reviewing the dynamics of vortices in the absence of a trap. The moduli space ${\cal M}_{N,k}$ is defined to be the space of solutions to the $U(N)$ vortex equations \eqn{vortex} with magnetic charge $k$. Index theorems show that \cite{erick,vib}
\be {\rm dim}({\cal M}_{N,k}) = 2Nk\nn\ee
The low-energy dynamics of vortices can be described by motion on ${\cal M}_{N,k}$ \cite{manmon}.  As we now review, the kinetic terms in \eqn{lag} induce a natural metric on ${\cal M}_{N,k}$ such that the path of followed by vortices as they scatter is a geodesic on ${\cal M}_{N,k}$.

\para
We start by introducing coordinates $X^a$, $a=1,\ldots, 2Nk$ on ${\cal M}_{N,k}$. These are the collective coordinates of the vortices, meaning that the most general vortex solution can be written as  $A_z = A_z(x;X)$ and $q_i= q_i(x;X)$. We define the zero modes to be 
\be \delta_a A_z = \frac{\partial A_z}{\partial X^a} + {\cal D}_z \Omega_a\ \ \ {\rm and}\ \ \ \delta_a q_i = \frac{\partial q_i}{\partial X^a} + i\Omega_a q_i\nn\ee
where $\Omega_a(x;X)$ is a gauge transformation. By construction, these obey the linearised version of the vortex equations \eqn{vortex} for any choice of $\Omega_a$. To determine $\Omega_a$, we need to fix a gauge. We ask that the zero modes obey the following equations:
\be {\cal D}_z \delta_a q_i - i \delta_a A_z \,q_i = 0\ \ \ {\rm and}\ \ \ {\cal D}_{\bar{z}}\delta_a A_z = -\frac{ie^2}{2}\delta_a q_i\,q_i^\dagger\label{linearised}\ee
The first of these is the linearised version of the first vortex equation in \eqn{vortex}; it is obeyed for any choice of $\Omega$. The second equation combines the linearised second vortex equation in \eqn{vortex} with a gauge fixing condition; it determines $\Omega_a$.  

\para
The key idea of the moduli space approximation is that we can accurately describe vortex dynamics by simply promoting $X^a\rightarrow X^a(t,\sigma)$ where $\sigma =x^3$ parameterises the spatial extent of the string. If we further set $A_0 = -\Omega_a \dot{X}^a$ and $A_3 = -\Omega_a {X^{a}}^{\prime}$, then the kinetic terms for fields are given in terms of the zero modes
\be {\cal D}_\alpha q_i  = \delta_a q_i\,\partial_\alpha X^a\ \ \ {\rm and}\ \ \ F_{\alpha i} = \delta_aA_i\,\partial_\alpha{X}^a\ \ \ \ \alpha=0,3\nn\ee
The gauge fixing condition in \eqn{linearised} has been chosen so that these fields satisfy the Gauss' law constraints from the gauge theory.

\para
Substituting this ansatz into the kinetic terms in \eqn{lag} gives us our expression for the vortex dynamics in terms of a non-linear sigma-model on ${\cal M}_{N,k}$,
\be S_{\rm vortex} = \int dtd\sigma \ g_{ab}(X)\,\partial_\alpha X^a\,\partial^\alpha X^b\nn\ee
where the metric $g_{ab}$ is given by
\be g_{ab} =  \int d^2x\ \frac{2}{e^2}{\rm Tr}\,(\delta_a A_z \delta_b A_{\bar{z}} + \delta_a A_{\bar{z}}\delta_b A_z)  + \sum_{i=1}^N(\delta_a q_i^\dagger \delta_b q_i + \delta_b q_i^\dagger \delta_a q_i)\nn\ee

\subsubsection*{Adding the Harmonic Trap}

Now let's see how things change when we add the harmonic trap. In general, this increases the energy of a vortex solution. When evaluated on the vortex equations \eqn{vortex}, the potential terms in \eqn{lag} simplify and can be written as
\be V =  \int d^2x\ \frac{2}{e^2} {\rm Tr}\,|{\cal D}_z\phi - \frac{\omega}{2}\bar{z}B_3|^2 + \sum_{i=1}^N |\phi q_i - \omega \bar{z}{\cal D}_{\bar{z}}q_i|^2\label{potv}\ee
The potential vanishes only if equations \eqn{vphi} are obeyed. These are the surviving, static vortices in the presence of a trap. For other vortex configurations, we should determine $\phi$ by minimising $V$ in the vortex background.  In this manner, the potential  $V$ can be thought of a  potential over the moduli space ${\cal M}_{N,k}$.  

\para
It is useful to have an expression for the potential $V$ purely in terms of moduli space data. We will prove the following
\be   V = g_{ab}\, K_{\rm rot}^a K_{\rm rot}^b\label{claim}\ee
where $K_{\rm rot}$ is the Killing vector on ${\cal M}_{N,k}$ arising from spatial rotations in the $(x^1,x^2)$ plane. 

\para
To show this, we first need to examine more closely the zero modes associated to spatial rotations.  They take the form
\be \delta_{\rm rot} q_i &=& i\omega (z\partial_z - \bar{z}\partial_{\bar{z}}) q_i + i \Omega q_i \nn\\ \delta_{\rm rot} A_z &=& \omega(z\partial_z - \bar{z}\partial_{\bar{z}})A_z + i A_z + {\cal D}_z\Omega\nn\ee
where the extra term in $\delta_{\rm rot}A_z$ reflects the fact that the field has spin 1. We've chosen to normalise the zero modes by $\omega$, the strength of the trap. 
To make progress, we need to identify the  gauge transformation $\Omega$. We will see that it is useful to decompose the gauge transformation as 
\be \Omega = -i\omega(zA_z - \bar{z}A_{\bar{z}}) + \hat{\phi}\nn\ee
Note that we've elected to call this gauge transformation $\hat{\phi}$. This is not a coincidence. With this choice the rotational zero modes can be conveniently written as 
\be 
\delta_{\rm rot} q_i &=& -i\bar{z\omega}{\cal D}_{\bar{z}} q_i + i\hat{\phi} q_i \nn\\ \delta_{\rm rot} A_z  &=& - \frac{\bar{z}\omega}{2} B_3 + {\cal D}_z\hat{\phi}\label{rotzero}\ee
It remains to determine $\hat{\phi}$. As we have described, this is fixed by the second of the equations in \eqn{linearised} which becomes
\be {\cal D}_{\bar {z}}{\cal D}_z \hat{\phi} = \frac{\omega}{2}B_3 + \frac{e^2}{2}\hat{\phi} q_i q_i^\dagger\nn\ee
But this is rather nice. It is precisely the equation obeyed by the field $\phi$ that arises from minimising \eqn{potv} in the background of a vortex. In other words,  $\hat{\phi} = \phi$. 
In particular, this highlights a claim that we made  previously: for rotationally invariant vortices, the field $\phi$ can be thought of as a compensating gauge transformation.

\para
The overlap of the rotational zero modes gives the length squared of the Killing vectors. From \eqn{rotzero}, we see that it can be written as
\be g_{ab} K^a_{\rm rot}K^b_{\rm rot} = \int d^2x\ \frac{2}{e^2} {\rm Tr}\,|{\cal D}_z\phi - \frac{\omega}{2}\bar{z}B_3|^2 + \sum_{i=1}^N |\phi q_i - \omega \bar{z}{\cal D}_{\bar{z}}q_i|^2\nn\ee
which proves our claim \eqn{claim}. 
\para
In the presence of a harmonic trap, there are also new contributions to the kinetic terms, coming from ${\cal D}_0\phi $ terms in \eqn{lag}. These are of order ${\cal O}(\omega^2\dot{X}^2)$; they can be ignored only if we think of both derivatives and the potential being small. If this is the case, the moduli space approximation for the dynamics in the presence of the trap is given by
\be S_{\rm vortex} = \int dtd\sigma\ g_{ab}(X)\,\partial_\alpha X^a\partial^\alpha X^b -  g_{ab}(X)K_{\rm rot}^a K_{\rm rot}^b\nn\ee

\section{Appendix: Dynamics of $U(1)$ Vortices in a Trap}

In this appendix, we describe the dynamics of $U(1)$ vortices in a harmonic trap. For simplicity, we focus on vortex particles in $d=2+1$ dimensions rather than the vortex strings in $d=3+1$ described elsewhere in this paper. 

\para
Each vortex is a disc of magnetic field of radius $\sim 1/ev$. As $k$ vortices are brought together, they form a larger disc of radius $R \approx\sqrt{k}/ev$. This fact has motivated the bag model of vortices \cite{stefano1,stefano2} in which vortices are thought of as an incompressible fluid. 
However, applications of the bag model for BPS vortices have  always been limited by the fact that vortices could fragment at zero cost of energy. 

\para
The presence of a harmonic trap prohibits this fragmentation. Now we can ask what the low-energy dynamics of BPS vortices looks like. We will show that, as expected from the bag model, the low-energy dynamics of $k$ vortices can be described by a free scalar field propagating around the edge of the disc. 

\para
We work with the matrix model of vortex dynamics described in Section \ref{kinksec}. (It would be interesting to revisit this calculation using a more direct approach, perhaps using the coordinates on the vortex moduli space introduced in \cite{samols}.) As we have seen, the unique ground state of $k$ vortices is given by \eqn{gs}. Here we are interested in excitations about this ground state which obey the linearisation of the constraint \eqn{vconstraint}. These were described in \cite{alexios}. They take the particularly simple form,
\be \delta_l Z = (Z_k^\dagger)^{l-1} \ \ \ {\rm and}\ \ \ \ \delta_l\varphi = 0 \ \ \  \ {\rm with}\ \ \ \ l=1,\ldots,k\label{pert}\ee
There are no higher excitations because $(Z_k^\dagger)^k=0$. We then describe the most general excitation as
\be Z(t)  = Z_0 + \sum_{l=1}^{n} c_l(t) Z_0^{\dagger\,l-1}\nn\ee
with the complex coefficients $c_l$ providing coordinates on the moduli space in the neighbourhood of the ground state. We work in a gauge with vanishing $U(k)$ gauge field. Then the kinetic terms in \eqn{vaction} are
\be S_{\rm kin} = \int dt\ {\rm Tr}\,|{\cal D}_tZ|^2 + |{\cal D}_t\varphi|^2 = \int dt\ ({\rm Tr}\, Z_k^{l-1} Z_k^{\dagger\,n-1} )\,\dot{c}_l^\star\dot{c}_n\nn\ee
Meanwhile, the potential term from \eqn{vaction} is 
\be S_{\rm pot} = -\int dt\ {\rm Tr}\,|[\sigma,Z]+ \omega Z|^2 + \varphi^\dagger \sigma^2 \varphi = -\int dt\ ({\rm Tr}\, Z_k^{l-1} Z_k^{\dagger\,n-1} )\,\omega^2 l^2 c_l^\star c_n\nn\ee
where we've used \eqn{wzero} and the expression for $\sigma$ given below \eqn{gs}.  We're left having to compute the traces
\be {\rm Tr}\,Z_k^{l-1}Z_k^{\dagger\,n-1} \equiv \Theta_l\delta_{ln}\ \ \ {\rm with}\ \ \ \Theta_l= \frac{r^{l-1}}{l}k(k-1)\ldots(k-l+1)\nn\ee
The low-energy dynamics of the vortices can then be written as
\be S = \sum_{l=1}^k\Theta_l \int dt\ \left(\dot{c}_l^{\,\star}\,\dot{{c}}_l - \omega^2 l^2 {c}_l^{\,\star}{c}_l\right)\nn\ee
This can be interpreted as the action for a real, massless scalar field, propagating around the edge of the vortex disc, truncated to the lowest $k$ Fourier modes. The zero mode is also projected out. If the disc has radius $R\approx \sqrt{k}/ev$, the scalar field has velocity $v=\omega R$. As $k$ increases, the radius of the disc scales as $\sqrt{k}$, which means that the density of modes scales as $1/\sqrt{k}$. This suggests that there is a continuum $d=1+1$ dimensional limit as $k\rightarrow \infty$.

\para
The calculation above closely follows an analogous calculation for vortices in a first order system \cite{carlme}. The difference is that the edge modes for relativistic vortices propagate both ways around the circle while the non-relativistic vortices described in \cite{carlme} have only chiral excitations.

\para
Here we have described the dynamics of $U(1)$ vortices in a trap. The dynamics of $U(N)$ vortices seems somewhat harder. In the presence of masses $m_i$, there are isolated ground states for the vortices, with the confined monopoles  now acting as instantons interpolating between these vacua. It would be interesting to understand if there is a corresponding boundary interpretation for the resulting dynamics.

\section{Appendix: An Index Theorem for Non-Abelian Kinks}\label{indexapp}

In this appendix, we derive the result of the index theorem \eqn{index}. 

\para
The general path of the calculation follows previous index theorems for kinks in gauge theories. The index theorem for kinks in Abelian gauge theories was first derived in \cite{lee} and for non-Abelian gauge theories, in the absence of the adjoint scalar $Z$, in \cite{sakaime}. Both of these, in turn, closely follow the index theorem of E. Weinberg for magnetic monopoles \cite{ejw1,ejw2}. As we will see, the presence of the adjoint scalar $Z$ makes the computation somewhat more involved.

\para
We start by rewriting the kink equations \eqn{kink} in slightly more useful notation which allows us to avoid  keeping track of the flavour indices $i$. We  define a $k\times N$ matrix $\varphi$ whose columns are $\varphi_i$, and also a mass matrix $m = {\rm diag}(m_1, \ldots, m_N)$. Then we can write the kink equations in the more convenient form
\be {\cal D}_3 \varphi &=& (\sigma - m_R) \varphi \nn \\
{\cal D}_3 Z &=& [\sigma,Z] +\omega Z \label{kinkagain} \\
{\cal D}_3 \sigma &=& g^2 ([Z,Z^\dagger] + \varphi \varphi^\dagger - r)  \nn \ee
where the subscript $R$  denotes how the operator acts: $m_R \varphi \equiv \varphi m$. We count the number of zero modes of these equations, which means that we count the number of solutions to the linearised kink equations. Given a background solution to \eqn{kinkagain} we write perturbations as hatted variables so, for example, $\varphi \to \varphi + \hat{\varphi}$. The linearised equations are then
%
%
\be \partial_3 \hat{\varphi} &=& (\hat{\sigma} + i \hat{\alpha}_3)\varphi + (\sigma + i \alpha_3  - m_R) \hat{\varphi} \nn \\
\partial_3 \hat{Z} &=& [\hat{\sigma} + i \hat{\alpha}_3,Z] + [\sigma + i \alpha_3,\hat{Z}] +\omega \hat{Z} \nn \\
\partial_3 \hat{\sigma} &=& i [\alpha_3, \hat{\sigma}] + i [\hat{\alpha}_3, \sigma] + g^2 ([\hat{Z},Z^\dagger] + [Z,\hat{Z}^\dagger] + \hat{\varphi} \varphi^\dagger + \varphi \hat{\varphi}^\dagger)  \nn \ee
To this we add a background gauge fixing condition,
\be \partial_3 \hat{\alpha}_3 =  i[\sigma, \hat{\sigma}] + i[\alpha_3, \hat{\alpha}_3] -ig^2( [\hat{Z},Z^\dagger] - [Z,\hat{Z}^\dagger] + \hat{\varphi} \varphi^\dagger - \varphi \hat{\varphi}^\dagger ) \nn \ee
This naturally combines with the third linearised  equation. We defined the complex-valued adjoint $U(k)$ object  $\xi = \sigma + i \alpha_3$, where $\alpha_3$ is the spatial component of the $U(k)$ gauge field and these can be written in the unified form 
\be \partial_3 \hat{\xi} &=& [\hat{\xi}, \xi^\dagger] + 2g^2 ([\hat{Z},Z^\dagger]  + \hat{\varphi} \varphi^\dagger )  \nn\ee
The number of solutions to these equations should be independent of $g^2$; to simplify the calculation we set $2g^2=1$. We then write the adjoint of these equations as
\be \Delta  \left(\begin{array}{c} \hat{\xi}^\dagger \\ \hat{\varphi}^\dagger \\ \hat{Z}^\dagger \end{array}\right) = 0 \quad \mbox{,} \hspace{5em} \Delta \equiv \left(\begin{array}{ccc}\partial_3 - \xi_a  & -\phi_L & -Z_a \\ -\phi^\dagger_L & \partial-\xi^\dagger_R + m & \\ - Z^\dagger_a & & \partial + \xi^\dagger_a -\omega \end{array}\right)\nn \ee
Once again, subscripts denote the action of the operator: $a$ means adjoint, with  $\xi_a X = [\xi, X]$; $L$ means left multiplication, with  $\phi^\dagger_L X = \phi^\dagger X$. Our goal is to find the number of solutions to these equations or, equivalently, the number of zero eigenvalues of $\Delta$.

\para
We can also construct the adjoint operator to $\Delta$, namely
\be \Delta^\dagger = \left(\begin{array}{ccc}-\partial^\dagger_3 - \xi_a  & -\phi_L & -Z_a \\ -\phi^\dagger_L & -\partial-\xi_R + m & \\ - Z^\dagger_a & & -\partial + \xi_a - \omega \end{array}\right) \nn\ee
Here there is a very important point: it is {\it not} true that $\Delta^\dagger$ has no zero modes. This is in contrast to most index theorems for BPS solitons. It means that, for some range of parameters, the index that we compute undercounts the number of zero modes of the configuration; we give some examples in the main text.

\para
We define the regularized index 
\be {\cal I}\left(M^{2}\right)=\mathrm{Tr}\left[\frac{M^{2}}{\Delta^{\dagger}\Delta+M^{2}}-\frac{M^{2}}{\Delta\Delta^{\dagger}+M^{2}}\right] \nn\ee
where the trace is a functional one. The index that we want is
\be {\cal I}  \equiv \lim_{M\to 0} {\cal I}(M^2) \nn\ee
This counts the number of zero modes. 

\para
A standard set of manipulations \cite{lee,sakaime,ejw1,ejw2} allows us to express the index in terms of a total divergence of a current, and hence as a boundary term,
\be {\cal I}(M^2) = \frac{1}{2}\left[J(x^3,M^{2})\right]_{x^3=-\infty}^{x^3=+\infty} \nn\ee
where
\be J(x^3,M^{2}) = \mathrm{Tr}\left\langle x^3\,\big|\, j \,\big| \,x^3\right\rangle  \label{indexcurrent} \ee
and 
\be 
 j = \Delta\frac{1}{\Delta^{\dagger}\Delta+M^{2}}+\Delta^{\dagger}\frac{1}{\Delta\Delta^{\dagger}+M^{2}}\nn \ee
Hence the problem reduces to computing this trace in this particular limit. Conveniently, we only need the trace in the asymptotic regime where $(\xi, Z, \phi)$ solve the vortex equations \eqn{vconstraint} and \eqn{wzero}. In particular, we can use $\xi = \xi^\dagger = \sigma$. Asymptotically, the two squared operators are identical and take the simple form
\be \left.\Delta\Delta^\dagger\right|_{x^3=\pm \infty}= \left.\Delta^\dagger\Delta\right|_{x^3=\pm \infty} = \left(\begin{array}{ccc} \Theta_1  & &  \\ & \Theta_2 & \varphi^\dagger_L Z_a \\ & Z^\dagger_a\varphi_L & \Theta_3 \end{array}\right)\label{almosteasy} \ee
where
\be \Theta_1 &=& -\partial^{2}+\sigma_{a}^{2}+\varphi_{L}\varphi_{L}^{\dagger}+Z_{a}Z_{a}^{\dagger} \nn \\
\Theta_2 &=& -\partial^{2}+(\sigma_{R}-m)^{2}+\varphi_{L}^{\dagger}\varphi_{L} \nn \\
\Theta_3 &=& -\partial^{2}+(\sigma_{a}-\omega)^{2}+Z_{a}^{\dagger}Z_{a} \nn \ee
Note, however, that the presence of the adjoint scalar $Z$ means that the matrix \eqn{almosteasy} is not block-diagonal. This makes the computation mathematically trickier than that of \cite{sakaime}, but ultimately these terms will not contribute to the index. Instead, the possibility of fluctuations in Z are responsible for the
richer structure of the index theorem

\para
The traces in \eqn{indexcurrent} separate into two terms,
\be {\rm Tr}\, j &=& -2\,\mathrm{Tr}\, \sigma_{a}\left(\Theta_{1}+M^{2}\right)^{-1} +2\, \mathrm{Tr}  \left(\begin{array}{ccc}-\sigma_R + m & \\ & \sigma_a - \omega \end{array}\right) \left(\begin{array}{ccc} \Theta_2 + M^2 & \varphi^\dagger_L Z_a \\ Z_a^\dagger\varphi_L & \Theta_3 + M^2 \end{array}\right)^{-1} \nn \ee
To make progress, we need to use a number of properties of these matrices when evaluated on the vortex solutions \eqn{vacua}. We begin with the first term. One can show that $[\sigma_a, \varphi_L\varphi_L^\dagger + Z_a Z_a^\dagger] = 0$. This means that $[\sigma_a,\Theta_1]=0$, and so we may simultaneously diagonalise these two expressions. Denote the eigenvalues of $\sigma_a$ by $\lambda_a$, and the eigenvalues  of $\varphi_L\varphi_L^\dagger + Z_a Z_a^\dagger$ by $\lambda'_a$. Then
\be \mathrm{Tr}\, \sigma_{a}\left(\Theta_{1}+M^{2}\right)^{-1} = \sum_a \frac{\lambda_a}{-\partial^2 + M^2 + \lambda_a^2 + \lambda'_a} \label{vanishingindexterm} \ee
A straightforward computation then  shows that
\be \sigma_a X = \lambda_a X \quad &\implies& \quad \sigma_a X^\dagger = -\lambda_a X^\dagger \nn \\
(\varphi_L\varphi_L^\dagger + Z_a Z_a^\dagger)X = \lambda'_a X \quad &\implies& \quad (\varphi_L\varphi_L^\dagger + Z_a Z_a^\dagger)X^\dagger = \lambda'_a X^\dagger \nn \ee
As a result, we find that all terms in \eqn{vanishingindexterm} pair up and annihilate each other. The first term in ${\rm Tr}\,j$ vanishes.

\para
We can treat the second term in ${\rm Tr}\,j$ similarly. We observe that
\be \left(\begin{array}{ccc} \Theta_2 + M^2 & \varphi^\dagger_L Z_a \\ Z_a^\dagger\varphi_L & \Theta_3 + M^2 \end{array}\right) = -\partial^2 + M^2 + Q^2 + R\nn \ee
where
\be Q = \left(\begin{array}{ccc}-\sigma_R + m & \\ & \sigma_a - \omega \end{array}\right)\ , \qquad R = \left(\begin{array}{ccc} \varphi^\dagger_L \varphi_L & \varphi^\dagger_L Z_a \\ Z_a^\dagger\varphi_L & Z_a^\dagger Z_a \end{array}\right)\nn  \ee
satisfy $[Q,R] = 0$. Denote the simultaneous eigenvalues of $Q,R$ as $\mu_A$ and $\mu'_A$. 
The number of solutions must be independent of simultaneously rescaling $(\varphi, Z, \sqrt{r})$ and so cannot depend on the  values of the $\mu'_A$. Technically this arises because, for each non-zero eigenvalue of $R$, the corresponding eigenspace has vanishing $Q$ trace. This means that we can process by setting $\mu_A'=0$ to arrive at the following expression:
\be J = \int \frac{dp}{2\pi} \,  \left\langle p \, \Big| \mathrm{Tr}\, j \Big| p \right\rangle =\sum_A \frac{\mu_A}{\sqrt{M^2 + \mu_A^2 }} \longrightarrow \sum_A {\rm sgn} \, (\mu_A)\nn \ee
We're left having to compute the signs of the eigenvalues of  $-\sigma_{R}+m$ and $\sigma_{a}-\omega$. Because $\sigma$ takes the diagonal form \eqn{vacua}, the eigenvalues of $-\sigma_R+m$ are simple:
\be
\sum_{i}\left(m_{j}-m_{i}+\left(q-1\right)\omega\right)\nn\ee
The eigenvalues of $\sigma_a-\omega$ are the differences of all pairs of diagonal elements of $\sigma$ (because $\sigma_a$ is in the adjoint representation). This gives
\be \sum_{i}\sum_{j}\sum_{q=1}^{k_{i}}\sum_{p=1}^{k_{j}}\mathrm{sg
n}\left(m_{j}-m_{i}+\left(q-p-1\right)\omega\right)
\nn\ee
Combining the two, we have our final expression for the index
\be {\cal I}   &=& \frac{1}{2} \left[ \sum_{i,j}
\sum_{q=0}^{k_i-1} \sum_{p = 0}^{k_j}
{\rm sign}(m_j - m_i + (q-p) \omega)\right]_{x^3=-\infty}^{x^3=+\infty}\nn\\ \  && \nn\\ & =& \frac{1}{2} \left[ \sum_{i,j}
\sum_{q=0}^{k_i-1}
{\rm sign}(m_j - m_i + (q-k_j)
\omega)\right]_{x^3=-\infty}^{x^3=+\infty}\nn\ee
where we've retained the first, more complicated version of the sum to
illustrate the symmetry in the counting.  By making use of the anti-symmetry in $(i,j)$ we arrive at the second line. This is the result \eqn{index} in the main text.

%

\end{document}